\documentclass[10pt,letterpaper]{article}
\usepackage{opex3}

\usepackage{amsmath,amssymb,graphicx}
\usepackage{epstopdf}

\usepackage{cite}

\usepackage{fixltx2e}

\begin{document}
\title{Identifying modes of large whispering-gallery mode resonators from the spectrum and emission pattern}



%

\author{Gerhard Schunk,$^{1,2,3,*}$ Josef U. F\"{u}rst,$^{1,2}$ Michael F\"{o}rtsch,$^{1,2,3}$ Dmitry V. Strekalov,$^1$ Ulrich Vogl,$^{1,2}$ Florian Sedlmeir,$^{1,2,3}$ Harald G. L. Schwefel,$^{1,2}$ Gerd Leuchs,$^{1,2,4}$ and Christoph Marquardt$^{1,2,5}$}
\address{
$^1$Max Planck Institute for the Science of Light, G\"{u}nther-Scharowsky-Stra\ss e 1/Building 24,\\ 90158 Erlangen, Germany\\
$^2$Institute for Optics, Information and Photonics, University Erlangen-N\"{u}rnberg, Staudtstr.7/B2, 90158 Erlangen, Germany\\
$^3$SAOT, School in Advanced Optical Technologies, Paul-Gordan-Str. 6,\\ 91052 Erlangen, Germany\\
$^4$Department of Physics, University of Ottawa, Canada\\
$^5$Department of Physics, Technical University of Denmark, Building 309,\\ 2800 Lyngby, Denmark
}
\email{$^*$Gerhard.Schunk@mpl.mpg.de}
\begin{abstract}
Identifying the mode numbers in whispering-gallery mode resonators (WGMRs) is important for tailoring them to experimental needs. Here we report on a novel experimental mode analysis technique based on the combination of frequency analysis and far-field imaging for high mode numbers of large WGMRs. The radial mode numbers q and the angular mode numbers p\,=\,$\ell$-m are identified and labeled via far-field imaging. The polar mode numbers $\ell$ are determined unambiguously by fitting the frequency differences between individual whispering gallery modes (WGMs). This allows for the accurate determination of the geometry and the refractive index at different temperatures of the WGMR. For future applications in classical and quantum optics, this mode analysis enables one to control the narrow-band phase-matching conditions in nonlinear processes such as second-harmonic generation or parametric down-conversion.
\end{abstract}
%
%
%

%
\section*{Introduction}
The outstanding properties of whispering-gallery mode resonators (WGMRs) \cite{Matsko2006,Braginsky1989a} have enabled breakthroughs in various fields such as single-particle sensing \cite{Vollmer2008,Sedlmeir2014}, coupling to single atoms \cite{Vernooy1998,Aoki2006a}, narrow-band optical filtering \cite{Vyatchanin1992} and lasing \cite{Sprenger2009}, optomechanical effects \cite{Hofer2010,Herr}, and WGMR-enhanced nonlinear optics \cite{Yu1999,Savchenkov2004,Furst2010natural,Furst2010,OPN,Strekalov2013a} being the basis for frequency comb generation \cite{Del'Haye2007,
Savchenkov2011} and quantum optics \cite{Furst2011,Michael2013}. 

WGMRs support a discrete set of eigenmodes, the so called whispering-gallery modes (WGMs). Most WGMR experiments could benefit from the knowledge of mode numbers for an exact quantification of their parameters. In sensing experiments with WGMRs, for instance, exact information on the spatial overlap of the WGM and the probe particle is advantageous \cite{Vollmer2008}. In nonlinear optics, information on spectral properties, in addition to the spatial overlaps is needed to address and quantify conversion channels \cite{Furst2010,Werner2012}.

There are straight forward analytical \cite{Lam1992} and numerical methods to calculate the mode structure in WGMRs that are only slightly larger than the wavelength. Numerical studies in large WGMRs, however, pose a significant problem as even with today's computational possibilities full three-dimensional modeling is extremely difficult. A complete description for the far-field emission patterns from spheroidal WGMRs, as being experimentally investigated in this work, is still an open question in the field. In addition, the high number of experimentally accessible modes in large WGMRs makes mode analysis a very challenging task.

Several experimental methods for mode analysis have been studied in the past, such as near-field probing \cite{Knight1995,Mazzei2005,Savchenkov2005a,Lin2010,Keng2014}, far field imaging \cite{Gorodetsky1994,Savchenkov2007,Carmon2008,Dong2008,Sedlmeir2013} and investigating the spectral response of these resonators \cite{Lin2010,Schiller1991,Preu2008,Li2012}. Regarding the breakthrough experiments mentioned earlier in the text, an experimental characterization of a macroscopic-size WGMR mode structure has only been achieved in the context of optical sum-frequency generation \cite{Strekalov2013a} on the basis of sideband spectroscopy \cite{Li2012}. This technique requires an optical probe coupled to the mode of interest at a wavelength that may not be available. Furthermore, finding the mode of interest, e.g. the signal or idler down-converted mode, with the probe laser is experimentally challenging.

The spatial structure and the resonance frequency of each WGM is characterized by a unique set of numbers, i.e. the polar, the azimuthal, and the radial mode number $\ell$, m, and q, respectively. The full information on these mode numbers is in principle contained in the far-field images of the outcoupled modes \cite{Gorodetsky1994}, and in the frequency spectrum \cite{Schiller1991}. The exact identification of  WGMs can be extremely difficult in practice using only one of these approaches. 

Here we present a combination of these techniques that allows for a complete identification of all significantly coupled WGMs of a large scale (millimeter-sized) WGMR. First, we determine the radial mode numbers q and the angular mode numbers p\,=\,$\ell$-m via an analysis of the observed far-field emission patterns. Then knowing these two numbers we unambiguously find the polar mode numbers $\ell$ by fitting the frequency spectrum. This fitting results in a very accurate determination of the product of the refractive index $n$ times the major radius $R$ of the resonator. Prior knowledge of either $n$ or $R$
allows the determination of the other parameter. This detailed characterization of the properties of the WGMR and the mode structure is important for WGMRs to become a versatile standard for classical and quantum optics.

The manuscript is structured as follows: section one theoretically describes far-field emission patterns based on the electric field distributions of WGMs. Section two explains the mode spectrum of WGMRs via the dispersion relation. After a description of the experimental setup in section three, experimental results on the characterization of the WGMR, in particular mode analysis, are presented on the basis of measured emission patterns and frequency spectra in section four.

\section{Theoretical background}

\subsection{Spatial analysis of whispering-gallery modes}\label{sec:spatana}
\begin{figure}[htb]
\centering
\includegraphics[width=.7\columnwidth]{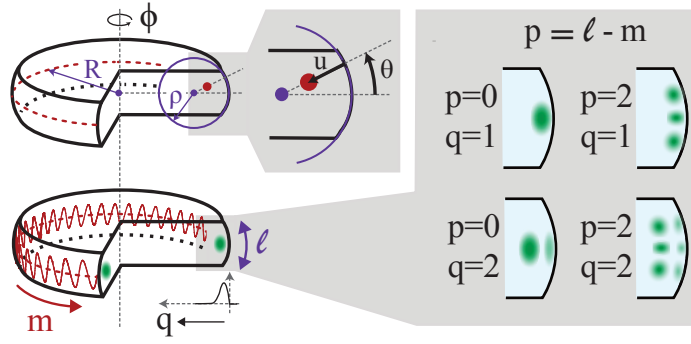}
\caption{Spatial structure of WGMs. The coordinate system for the electric field distributions defined by Eq.  \eqref{eq:spatwgmmain} is based on the radial distance $u$ of the observation point to the surface of the WGMR, the polar angle $\theta$, and the azimuthal angle $\phi$.  The intensity distributions on the right-hand side are shown for various radial mode numbers q and angular mode numbers  p=$\ell$-m. }
\label{fig:spatialWGMS}
\end{figure}

Far-field emission patterns from WGMRs carry information on the internal mode structure. According to Maxwell's equations, WGMs in spherically-symmetric resonators are described by spherical harmonic functions and spherical Bessel functions \cite{Oraevsky2002}. For WGMs close to the equatorial plane and the surface of a spheroidal WGMR, the electric field amplitudes are approximated by \cite{doi:10.1117/12.914606,Breunig2013a} (see Fig. \ref{fig:spatialWGMS}):
\begin{subequations}
	\begin{align}
		E \propto \underbrace{  e^{-\frac{1}{2}\left(\frac{\theta}{\theta_\textrm{m}} \right)^2 }
		\cdot H_\textrm{p}\left( \frac{\theta}{\theta_\textrm{m}} \right) \cdot e^{i \textrm{m} \phi} }_\textrm{angular part} 
		\cdot \underbrace{ \textrm{Ai}\left( \frac{u}{u_\textrm{m}} - \alpha_\textrm{q} \right)  }_\textrm{radial part} \,, \\
		\theta_\textrm{m} =\left( R/\rho\right)^{3/4} \textrm{m}^{-1/2} , \qquad   
		u_\textrm{m} = R/2^{1/3} \textrm{m}^{2/3} ,
		\label{eq:wgmrad}
	\end{align}
	\label{eq:spatwgmmain}
\end{subequations}
where $R$ is the major radius of the WGMR and $\rho$ the curvature. Analogous to the solution of Maxwell's equations in a box with Dirichlet boundary conditions, the WGMs in this spheroidal geometry are characterized by three integer mode numbers: the polar, the azimuthal, and the radial mode numbers $\ell\gg 1, \textrm{m} \gg 1$ and $\textrm{q}\geq 1$, respectively (see Fig. \ref{fig:spatialWGMS}). The angular mode number $\textrm{p}=\ell-\textrm{m}=0,1,2,...$ of the WGMs gives the degree of the Hermite polynomials and therefore the number of field oscillations in $\theta$-direction. $\theta_\textrm{m}$ is the angular size parameter. The radial part of the electric field in Eq.  \eqref{eq:wgmrad} is described by the radial size parameter $u_\textrm{m}$ and the q-th root $\alpha_\textrm{q}$ of the Airy-function $\textrm{Ai}(-\alpha$). 

The spatial structure of a WGM with open boundaries, where evanescent fields are present in the close vicinity of the WGMR, can be probed by placing a prism with refractive index $n_\textrm{prism}~>~n$ next to the WGMR. The far-field emission pattern from such a prism is described as a Fourier transform of the evanescent WGM field at the prism surface. This near-field at the prism surface can be derived by applying a coupling window \cite{Gorodetsky1994}, i.e. a finite aperture, to the angular part of Eq.  \eqref{eq:spatwgmmain}. This operation is equivalent to a low-pass filtering for the far-field. 

According to Eq. \eqref{eq:spatwgmmain}, the mode profile in $\theta$-direction contains information on the angular mode number p. For equatorial WGMs (p=0) (see Fig. \ref{fig:spatialWGMS}), the near-fields and therefore the far-fields show exactly one maximum. For $\textrm{p} > 0$, the far-fields in $\theta$-direction show two distinct lobes  symmetric around the equatorial plane for large spheres \cite{Gorodetsky1994}. This is not generally the case for aspherical resonators. For oblate resonators ($R > \rho$), these far-field distributions can also be modeled with two main maxima that are symmetric about this plane. According to the experimental results for our resonator ($R/\rho\approx3.8$) and p=2, these maxima overlap strongly (see section three for experimental results). For p=4, they can be resolved. For prolate resonators ($R<\rho$), the effect of the coupling window is reduced. In the extreme case of $R/\rho\ll1$, the width of the WGM in the near-field is much smaller than the width of the coupling window. Hence, the coupling window does not affect the out-coupling and more detailed features of the respective WGM mode structure are visible in the far-field.

For equatorial WGMs, the equatorial coupling angle $\Phi$ and the divergence $\Delta \Phi$ are given by \cite{Gorodetsky1999}:
\begin{align}
	\sin \Phi &=  \frac{\ell ~ \lambda}{n_\textrm{prism} ~2\pi R } \label{eq:equatoremm} \,, \\
	\Delta \Phi^2 &= \frac{\sqrt{n^2-1} ~ \lambda}{n_\textrm{prism}^2~2\pi R~\textnormal{cos}^2 \Phi} \hspace{0.1cm} \,.
\label{eq:equatordiv}
\end{align}
The equatorial angle $\Phi$ is in good approximation equivalent to the critical angle of total internal reflection of the resonator and the prism material and is independent of the minor radius $\rho$. A measurement of the equatorial angle $\Phi$ reveals the polar mode number $\ell$, which is equal to the azimuthal mode number m for equatorial modes.
\subsection{Spectral analysis of whispering-gallery modes}
An understanding of the mode spectrum of WGMRs requires a discussion of the dispersion relation \cite{Oraevsky2002,Breunig2013a,Gorodetsky2006} on the basis of frequency differences between individual WGMs. Using the dispersion relation \cite{Oraevsky2002,Gorodetsky2006}, we connect the mode numbers $\ell$, q, and $\textrm{p}=\ell-\textrm{m}$ to the resonant optical frequency $\nu_{\ell,\textrm{q,p}}$:
%
\begin{align}
 \nu_{\ell,\textnormal{q,p}} &= \underbrace{\frac{c}{2\pi n R} }_{1/x} \cdot \left[ \ell + \alpha_\textnormal{q} \left( \frac{\ell}{2}\right)^{{1}/{3}} + p \left( \sqrt{\frac{R}{\rho}} - 1 \right) -\frac{\chi \cdot n}{\sqrt{n^2-1}} + \frac{1}{2}\sqrt{\frac{R}{\rho}} \right. \nonumber \\
 &\quad+ \left. \frac{3 \alpha^2_\textnormal{q}}{20} \left(\frac{\ell}{2}\right)^{-{1}/{3}} + O\left( \ell^{-{2}/{3}} \right) \right]  ,
\label{eq:disprel}
\end{align}
%
where p and $\sqrt{R/\rho}$ are of order one. Depending on the polarization, the parameter $\chi$ is 1 for TE modes and $1/n^2$ for TM modes. The wavelength-dependent refractive index of the WGMR is $n$. This renders the right-hand side of Eq.  \eqref{eq:disprel} frequency dependent.The q-th root of the Airy function $\alpha_\textrm{q}>0$ can be approximated as $\alpha_\textrm{q} = [3 \pi/2  (\textrm{q} - 1/4) ]^{2/3}$.The scaling factor $x=(2\pi n R)/c$ appears as a refractive index dependent and thus frequency-dependent normalization factor. The spectrum around a certain WGM at frequency $\nu_{\ell,\textnormal{q,p}}$ is found by evaluating frequency differences:
\begin{align}
	\Delta \nu_{\ell,\textnormal{q,p}}(\Delta\ell,\Delta \textrm{q},\Delta \textrm{p}) =  \nu_{\ell+\Delta\ell,\textrm{q}+\Delta\textrm{q},\textrm{p}+\Delta\textrm{p}} - \nu_{\ell,\textrm{q,p}} \,
	\label{eq:deltap}
\end{align}	
to other WGMs at frequencies $\nu_{\ell+\Delta\ell,\textrm{q}+\Delta\textrm{q},\textrm{p}+\Delta\textrm{p}}$. Due to the material dispersion, Eq.  \eqref{eq:disprel} is an implicit function of the frequency. Hence, the correct description of the frequency differences requires a full numerical approach.

For a better qualitative understanding of the relevant quantities for the formation of the frequency spectrum, we can give analytic expressions using first order approximations. For this, we define a dispersive scaling factor ${x_{\textrm{d}}}$ by including the slope of the refractive index $\frac{\partial n}{\partial \nu}$ at the frequency $\nu_{\ell,\textnormal{q,p}}$ as:
\begin{align}
 {x_{\textrm{d}}} = {x} \cdot  \left(1 + \frac{\nu_{\ell,\textnormal{q,p}}}{n} \left. \frac{\partial n}{\partial \nu}\right|_{\nu_{\ell,\textnormal{q,p}}} \right),
\end{align}
Small contribution of the polarization dependent term ${\chi\cdot n}/{\sqrt{n^2-1}}$ in Eq.  \eqref{eq:disprel} are omitted.

\begin{figure}
\centering
\centerline{\includegraphics[width=.8\columnwidth]{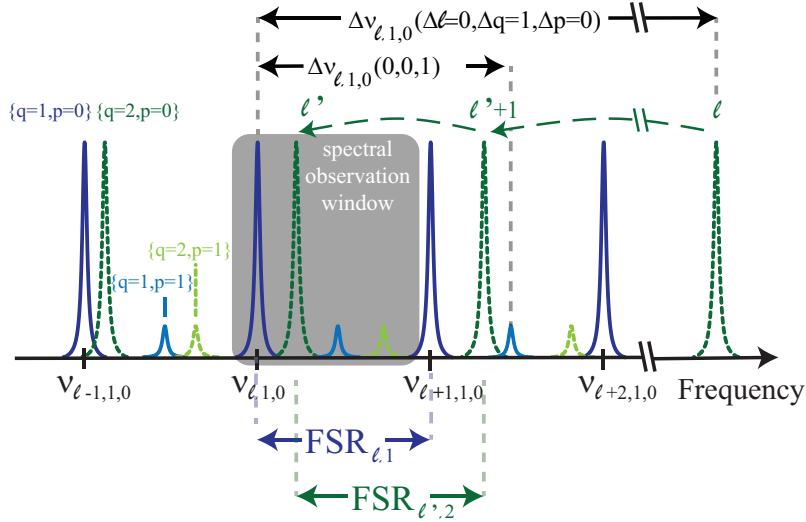}}
\caption{Illustration of a WGMR mode spectrum. As for all resonators, the free spectral range is the characteristic mode spacing for one mode family. Here, we show two mode families (dashed and solid lines), whose respective free spectral ranges ($\textnormal{FSR}_{\ell,\textrm{1}} \neq \textnormal{FSR}_{\ell ',\textrm{2}}$, see Eq.  \eqref{eq:fsrdip}) are determined by geometric dispersion. In addition to equatorial WGMs $\{\textrm{q}\geq 1,\textrm{p}=0\}$, each mode family contains non-equatorial WGMs $\{\textrm{q}\geq 1,\textrm{p}\neq 0\}$ at an offset frequency given by Eq.  \eqref{eq:diffp}. Within a given spectral observation window, the frequency differences of the WGMs belonging to different mode families allow to unambiguously identify the polar mode number $\ell$.
}
\label{fig:FSRmodespacing}
\end{figure}

In the following, we model the frequency spectrum on the basis of changes in the respective mode numbers $\ell,\textrm{q}$, and $\textrm{p}$. This discussion is illustrated in Fig. \ref{fig:FSRmodespacing}. The free spectral ranges $\textnormal{FSR}_{\ell,\textrm{q}}$ originate from steps in the $\ell$ number and depend only on $\ell$ and q:
\begin{align}
 \Delta\nu_{\ell,\textnormal{q,p}}(\Delta\ell=1,0,0) \equiv   \textnormal{FSR}_{\ell,\textrm{q}}
  \approx & \,	\frac{1}{x_{\textrm{d}}}  \left[ 1 + \frac{\alpha_\textrm{q}}{6}  \left( \frac{2}{\ell} \right)^{\frac{2}{3}} \right] . 
\label{eq:fsrdip}
\end{align}
For increasing polar mode numbers $\ell$, found for example at higher optical frequencies, WGMs are located closer to the WGMR surface. This increases the effective radius for these WGMs and decreases the $\textnormal{FSR}_{\textrm{q},\ell}$. In contrast, an increase in the radial mode number q leads to the field shifting away from the surface, which decreases the effective radius and increases the $\textnormal{FSR}_{\ell,\textrm{q}}$. 

The spacings between different radial WGMs:
\begin{align}
	\Delta\nu_{\ell,\textnormal{q,p}}(0,\Delta \textrm{q},0) 
	\approx \frac{1}{x_{\textrm{d}}} \left( \alpha_{\textrm{q}+\Delta\textnormal{q}} - \alpha_\textrm{q} \right) \left(\frac{\ell}{2}\right)^{\frac{1}{3}} 
	\label{eq:diffequ}
\end{align}
can exceed the $\textnormal{FSR}_{\ell,\textrm{q}}$ by orders of magnitude and depend only on $\ell$ and q. Within the spectral observation window, modes with a different radial number q will also have a different polar number $\ell'$ (see Fig. \ref{fig:FSRmodespacing}).

The spacings between WGMs with different angular mode numbers p:
\begin{align}
	\Delta\nu_{\ell,\textnormal{q,p}}(0,0,\Delta \textrm{p}) 
	\approx \frac{1}{x_{\textrm{d}}} \left( \sqrt{\frac{R}{\rho}} - 1 \right) \Delta \textrm{p} \,,
	\label{eq:diffp}
\end{align}
are determined solely by the ratio of the radii ${R/\rho}$  and independent on q and $\ell$. Eq.  \eqref{eq:diffp} has been used extensively in the context of frequency comb generation \cite{Savchenkov2011}. For $R/\rho\approx {N}^2$ where ${N} = 1,2,3...$, the spacings can match multiples of the $\textnormal{FSR}_{\ell,\textrm{q}}$ given by Eq.  \eqref{eq:fsrdip}, which can lead to degenerate frequencies $\nu_{\ell,\textnormal{q,p}} = \nu_{\ell+\textrm{p}\cdot(\textrm{N}-1),\textnormal{q,0}}$. The case of a spherical resonator (N=1), which exhibit degenerate mode families, is well known. According to Eq.~\eqref{eq:diffp}, spheroidal resonators may have sufficient symmetry to support degenerate mode families.

The frequency spectrum depends on a variety of parameters of the WGMR and the environment, such as temperature, material, pressure, and geometry. Each of these parameters can be used independently to tailor the frequency spectrum.

In principle, accurate knowledge about the scaling factor $x$, i.e. the fundamental value $n\cdot R$, together with an absolute frequency measurement of one WGM can provide a way to determine mode numbers $\ell$, q, and p of this particular WGM. This can be experimentally challenging due to the limited knowledge of the geometry and the refractive index of the WGMR. In contrast, the experimental study presented in the following two sections is based on relative frequency measurements. First, we obtain the mode numbers q and p from a spatial mode analysis. Using this knowledge, we  show a complementary analysis of the frequency differences $\Delta \nu_{\ell,\textnormal{1,0}} (\Delta\ell,\Delta \textrm{q},\Delta \textrm{p})$ from the fundamental WGM to higher-order WGMs to obtain the WGMR properties, such as the scaling factor $x$, the major radius $R$, and minor radius $\rho$. In this analysis, $\ell$ is exactly determined by a fitting procedure.

\section{Experimental setup}
\begin{figure}[htb]
\centerline{\includegraphics[width=0.7\columnwidth]{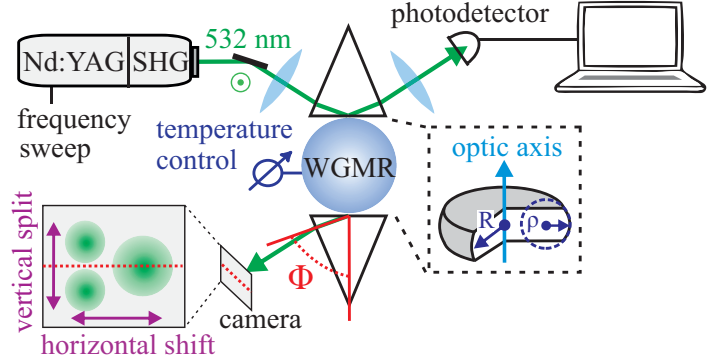}}
\caption{Illustration of the experimental setup showing the spectral and spatial mode analysis detection.}
\label{fig:setup}
\end{figure}

The experimental setup shown in Fig. \ref{fig:setup} allows spatial and spectral characterization of WGMs. The macroscopic WGMR is manufactured from a congruent 5.3\% MgO-doped lithium niobate wafer, such that the optic axis is aligned with the symmetry axis of the resonator (z-cut \cite{Furst2010}). The major and the minor radius of the disk are measured to be $R~=~1.594\pm0.006~$mm and $\rho~=~0.423\pm0.006~$mm with a microscope leading to a scaling factor of $x\approx 1/(13.4\,\textrm{GHz})$. The WGMR is mounted in an oven whose temperature is stabilized to a millikelvin level. The light source for probing the whispering-gallery resonator is a continuous wave laser at $532\,$nm (Nd:YAG Prometheus, Innolight) with a Hermite-Gauss TEM$_{00}$ mode. Its polarization is parallel to the optic axis of the WGMR (TE mode).

Evanescent coupling to the disk is achieved by focusing the beam onto the inner surface of a diamond prism under an angle of total internal reflection (see Fig. \ref{fig:setup}). The transverse coupling angle was adjusted by maximizing coupling to equatorial modes. All light coming from the coupling point is collimated and sent to a photo detector. At the resonance frequencies of individual WGMs, light can be coupled to the WGMR. Tuning the laser frequency over more than an FSR, this configuration can be used for the spectral analysis.

At the in-coupling prism, the emission pattern from the WGMR destructively interferes with the directly reflected spatial mode of the laser source. We use a second diamond prism for out-coupling to investigate the modes independently from this interference. The light emerging from the second prism is sent to a CCD camera placed at a distance of 1.4 cm from the coupling point, which gives us in a good approximation the far-field emission pattern.
 
\section{Experimental results}
In our mode analysis technique, we combine the information from far-field emission patterns and from the frequency spectrum. In the first subsection mode numbers q and p are obtained from an analysis of the far-field emission patterns. In the second subsection a complementary analysis of the frequency differences $\Delta \nu_{\ell,\textnormal{1,0}} (\Delta\ell,\Delta \textrm{q},\Delta \textrm{p})$ from the fundamental WGM to higher-order WGMs is performed to obtain the WGMR properties and the polar mode numbers $\ell$.

\subsection{Far-field imaging}
\begin{figure}[h!]
\centerline{\includegraphics[width=.78\columnwidth]{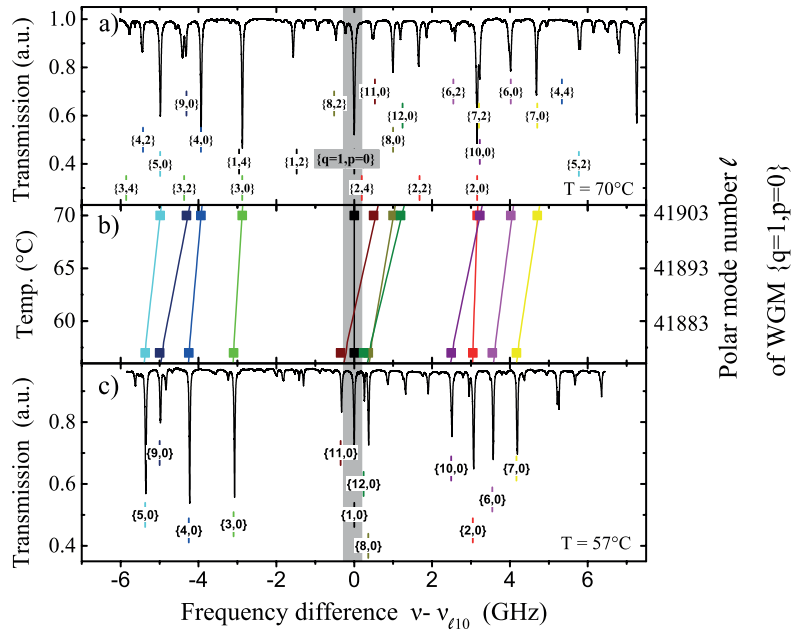}}
\caption{Labeled transmission spectra. Transmission spectra at $\lambda\,=\,532$\,nm of the macroscopic WGMR (major radius $R=1.59$\,mm) measured at $T=70\,^\circ$C (a) and $T=57\,^\circ$C (c). The radial mode numbers q and the angular mode numbers $\textrm{p}=\ell-\textrm{m}$ are assigned according to the far-field emission patterns (see Fig. \ref{fig:pics}). Frequencies are specified relative to the frequency $\nu_{\ell,\textrm{q=1,p=0}}$ of the fundamental WGM $\{\textrm{q=1,p=0}\}$, which is highlighted in grey. Non-equatorial WGMs $\{\textrm{q}\geq 1,\textrm{p}\neq 0\}$ exhibit a fixed frequency difference to the respective equatorial WGM $\{\textrm{q}\geq 1,\textrm{p=0}\}$ (see Fig. (a) and Eq. \eqref{eq:diffp}). Fig. (b) shows measured (square markers) and calculated (vertical lines) frequency differences $\Delta \nu_{\ell,\textnormal{1,0}}(\Delta\ell,\Delta \textrm{q},\Delta \textrm{p} = 0)$ according to Eq. \eqref{eq:deltap} for different equatorial WGMs $\{\textrm{q}\geq 1,\textrm{p=0}\}$ at different temperatures. Temperature tuning of these WGMs exhibits different slopes due to their different $\textrm{FSR}_{\ell,\textrm{q}}$ given by Eq.  \eqref{eq:fsrdip}. The frequency spectrum at a given temperature is then directly related to the polar mode number $\ell$ ($\ell=\textrm{m}$ for equatorial WGMs) of the fundamental WGM shown on the right axis.}
\label{fig:atlas}
\end{figure}

\begin{figure}[h]
\centering 
\includegraphics[width=.7\columnwidth]{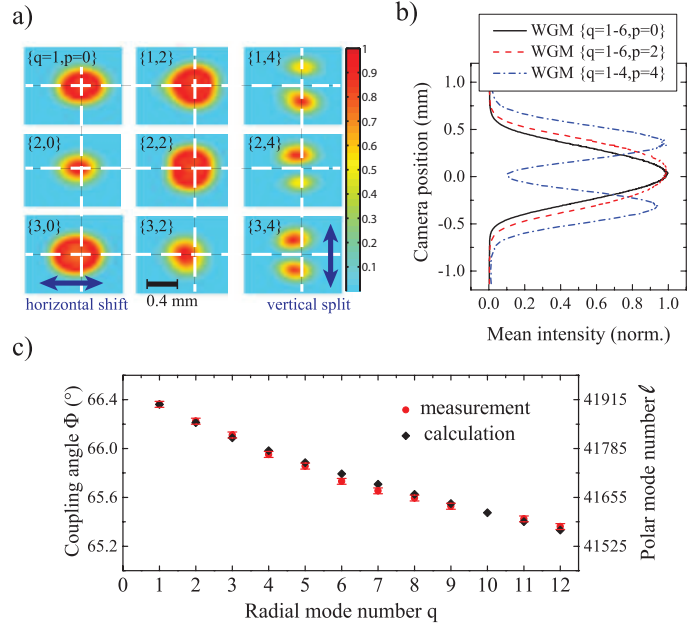}
\caption{a) Far-field emission patterns of various low-order WGMs (see Fig. \ref{fig:atlas}(a) for the corresponding frequency spectrum). The WGMs \{q=2,p=0\} and \{3,2\} exhibit distortions in the far-field images, which probably originate from their frequency degeneracy to other WGMs. The horizontal center positions of the intensity distributions shift to the left (towards smaller $\Phi$) for higher radial mode numbers q. b) Averaged vertical cross section for WGMs with the same p but different q. This gives information on the distribution of the WGMs in $\theta$-direction, which allows a distinction between different angular mode numbers p. c) Distinction of different radial mode numbers q of equatorial WGMs (p=0) according to the coupling angle $\Phi$. Within our spectral observation window, the higher the radial mode number q, the smaller is $\ell$ and $\Phi$. The exact polar mode numbers $\ell$ used for the calculated out-coupling angles $\Phi$ (see Eq.  \eqref{eq:equatoremm}) are derived from an analysis of the spectrum.}
   \label{fig:pics}
\label{fig:camerammfit}
\end{figure}

As a first step, we tune the pump laser frequency over 14~GHz and record the WGMR spectrum with a photodetector. It is known from sensing with WGMRs that objects within the evanescent fields, in this case the coupling prism, can induce a shift of the resonance frequencies \cite{Vollmer2008,Sedlmeir2014}. To avoid this effect, the experiment is carried out in the under-coupled regime of the fundamental WGM. The measured frequency spectrum is shown in Fig. \ref{fig:atlas}(a). The numbers $\{\textrm{q,p}\}$ of the modes shown in this figure are not yet determined. To find these numbers, we take far-field images (see Fig. \ref{fig:camerammfit}(a)) of all WGMs with a reasonably strong coupling within one FSR. With our coupling optimized for equatorial modes using a TEM$_{00}$ laser beam, the excitation of modes with odd p is strongly suppressed (see also \cite{Lin2010}). Consequently, we expect to see only even numbers of p, whose coupling decreases rapidly for higher p. Note that tilting the pump beam from the equatorial plane excites an increasing number of non-equatorial modes. The spectrum becomes more dense and less pronounced features of well-coupled modes are visible.

WGMs with different p numbers are distinguished based on the cross section of the far-field emission patterns, which is depicted in Fig. \ref{fig:camerammfit}(b) as an average for WGMs with the same p but different q. Equatorial WGMs with p=0 show a single-lobe nearly Gaussian emission pattern. WGMs with p=2 exhibit a flat-top angular distribution. This distinction is clear enough to identify the p=0 and p=2 modes in most relevant cases. For WGMs with p=4, a distinct two-lobe structure appears. A limitation arises for strongly oblate ($R \gg \rho$) resonators when emission patterns coming from p=0 and p=2 WGMs strongly overlap.

The measured emission patterns of the WGMs were fitted with Gaussian functions in horizontal direction. This reveals the information on the central positions (see in Fig. \ref{fig:camerammfit}(c) for the associated coupling angles $\Phi$) and $1/$e-values of the Gaussian functions, and hence the divergences of the emission patterns. The central position give the information on the q number. 

Since the resonance frequencies $\nu_{\ell,\textnormal{q,p}}$ of all WGMs within the spectral observation window (see Fig. \ref{fig:FSRmodespacing}) are approximately equal to the pump laser frequency, Eq. \eqref{eq:disprel} implies that larger radial mode numbers q correspond to smaller numbers $\ell$. Hence, an increase in the radial mode number q results in a decrease in the out-coupling angle $\Phi$ for equatorial WGMs according to Eq. \eqref{eq:equatoremm}. This corresponds to the horizontal shift of the center of the spatial profiles depicted in Fig. \ref{fig:pics}(a), which provides the basis for identification of q. This also means that coupling of modes with different q can be optimized via the coupling angle $\Phi$, which is observed in the experiment.

Therefore the analysis of the far-field emission patterns has allowed an assignment of radial mode number q and angular mode number p in the mode spectrum shown in Fig. \ref{fig:atlas}. The polar mode numbers $\ell$ are obtained from an analysis of frequency differences in the mode spectrum, which is discussed below. With the information on $\ell$, q, and p, one can calculate the coupling angles $\Phi$ using Eq.  \eqref{eq:equatoremm}. The calculated angles can be compared with the coupling angles inferred from geometry of the setup and the measured central positions of the far-field images. A constant offset angle is taken as a fitting parameter. The experimental results shown in Fig.~\ref{fig:camerammfit} are in good agreement with the model. Multiple measurements of one mode were taken to estimate the standard deviations for the measured coupling angles, which are presented as error bars in Fig. \ref{fig:pics}(c). For high radial mode numbers q, the error bars of adjacent WGMs will start to overlap. Adjacent WGMs can then no longer be distinguished unambiguously due to statistical variation in the measured central positions. Our assignment of radial mode numbers for equatorial WGMs up to q = 12 is also confirmed by the frequency analysis described below.

The experimental $1/$e-values of the mode profiles in $\Phi-$direction can now be translated into angular divergences. The mean value of the measured divergences $\Delta \Phi_\textrm{measured} = 0.60\,^\circ$ matches well (within $8\%$) with the theoretical value $\Delta \Phi_\textrm{theoretical} = 0.53\,^\circ$ (see Eq. \eqref{eq:equatordiv}). The divergences are approximately equal for all measured radial modes. 

\subsection{Spectral analysis}
We use our knowledge of the mode numbers q and p for the spectral analysis. The frequency of the fundamental WGM \{q=1, p=0\} is taken as a reference point. In Fig. \ref{fig:atlas}, all other modes were plotted relative to this fundamental WGM frequency. The measured frequency differences of the equatorial modes are fitted with calculated frequency differences according to the dispersion relation given by Eq. \eqref{eq:disprel}. For p=0, the dispersion relation depends on the radius $R$ and polar mode number $\ell$ as free parameters.
\begin{figure}[htb]
\centerline{\includegraphics[width=.8\columnwidth]{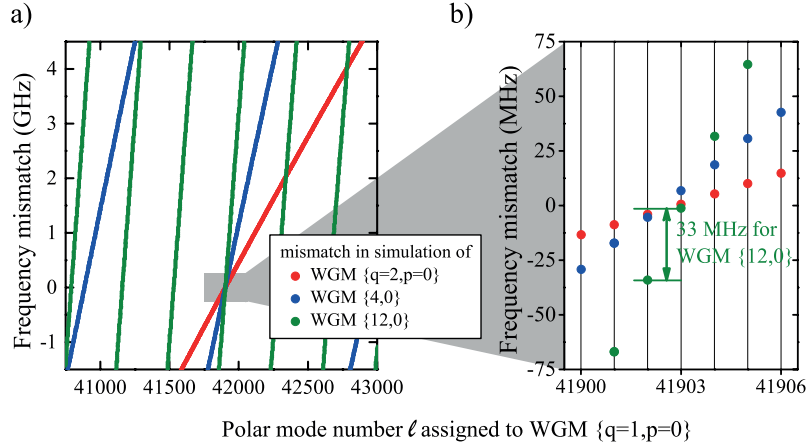}}
\caption{Evaluation of the measured frequency spectrum (see Fig. \ref{fig:atlas}(a)). a$)$ The mismatch between measured and computed frequency differences (see Eq.  \eqref{eq:deltap}) over a broad range of $\ell$ numbers of the fundamental WGM \{q=1,p=0\}. The frequency mismatches of a selection of equatorial WGMs are shown. The individual mismatches reach zero at multiple values of $\ell$, however, all together they reach zero at the unique value of $\ell$ = 41903 in b).}
\label{fig:accuracy}
\end{figure}

This fit (see Fig. \ref{fig:accuracy}) unambiguously yields the polar mode number $\ell$=41903 for the fundamental WGM, and confirms the assignment of q and p for the other WGMs gained from the spatial analysis. For p=0, the radius only appears in the scaling factor $x$ in the dispersion relation. Therefore the fitted radius $R$ can only be trusted within the accuracy for the absolute value for measured frequency differences. A much better accuracy is achieved by evaluating the fitted polar mode number together with absolute frequency of the pump laser. Taking the measured frequency $\nu_{\textnormal{$\ell$,1,0}}=(563.26 \pm 0.01)\,$THz of the fundamental WGM determined with a wavemeter, the fit results in an optical circumference of $n\cdot R=(3.55509 \pm 0.00006)\,\textrm{mm}$. 

Using the refractive index found from the Sellmeier equation for lithium niobate \cite{Schlarb1994}, $n=2.2244 \pm 0.0011$, we estimate the large radius to $R=(1.5982\pm0.0008)\,$mm at 70$\,^\circ$C. This is consistent with the value measured with the microscope but has a precision improved by an order of magnitude. Additionally, knowledge of the mode numbers and the frequency spacing between non-equatorial modes described by Eq.  \eqref{eq:deltap} can be used to fit the ratio $R/\rho$. Together with the knowledge of $R$, the fit yields $\rho=(0.4252 \pm 0.0011)\,$mm. The measurements of the radii also agree well with the microscopically measured value. Thermal expansion does not affect the WGMR radii within the accuracy of the microscope measurement.

In a final step, the spectral characterization is carried out again at a different temperature of the WGM resonator within the same spectral observation window (see Fig. \ref{fig:atlas}(c)). This results in a shift of the relative frequencies for all the modes within the spectrum (see Fig. \ref{fig:atlas}(b)). A change in temperature effectively changes the resonator optical length, both through thermal expansion and the thermal change in the refractive index of lithium niobate. This changes the size parameter $x$ of the resonator. The change in the measured mode spectrum in the fixed spectral observation window can thus be directly related to a change in the $\ell$ number for each mode in the observation window by $\Delta \ell=45$. In principle, temperature tuning can also be used to obtain accurate information on the thermo-refractive and the thermal expansion coefficient. Due to experimental uncertainties in the temperature determination, this is not possible here with a reasonable accuracy. In contrast, we use temperature tuning to demonstrate an efficient method for tailoring the frequency spectrum by choosing the proper $\ell$ number at a certain temperature. 
\section{Conclusions}
In summary, we have demonstrated a practical experimental technique to identify mode numbers of all significantly coupled WGMs within the spectral observation window by evaluating far-field emission patterns and frequency differences. This technique takes advantage of geometrical dispersion arising in WGMs due to their curvature. Therefore it is more efficient in application to small resonators. However, we have shown that the exact determination of the mode numbers is possible even in relatively large resonators.

Apart from analyzing an unknown mode spectrum, the understanding of the frequency spectra can also be used to specifically tailor WGM spectra. In addition, this measurement technique allows for accurate measurements of the resonator radii $R$ and $\rho$ times refractive index $n$. The determination of the optical circumference $n \cdot R$ is mainly limited by the uncertainty of the laser frequency. This can be increased experimentally by orders of magnitude to allow very accurate measurements of either resonator geometry or refractive index. The large set of accurately characterized WGMs can be employed to select WGMs with specific spatial or spectral properties on demand for a huge variety of applications in classical and quantum optics.

\section*{Acknowledgments}
The authors are grateful for the financial support of the European Research Council under the Advanced Grant PACART. Finally, the authours would like to thank Peter Banzer, Thomas Bauer and Martin Neugebauer for fruitful discussions. 
\end{document}